\documentclass[a4paper]{jpconf}
\usepackage{graphicx}

\newcommand{\sigmav}{$\langle\sigma v\rangle$ }

\begin{document}
\title{Isotropic diffuse and extragalactic $\gamma$-ray background: emission from extragalactic sources vs dark matter annihilating particles}

\author{Mattia Di Mauro, for the Fermi-LAT Collaboration}

\address{Kavli Institute for Particle Astrophysics and Cosmology, Department of Physics and SLAC National Accelerator Laboratory, Stanford University, USA. \\
E-mail: mdimauro@slac.stanford.edu}

\begin{abstract}
The isotropic diffuse $\gamma$-ray background (IGRB) has been detected by various experiments and recently the {\it Fermi}-LAT Collaboration has precisely measured its spectrum in a wide energy range.
The origin of the IGRB is still unclear and we show in this paper the significative improvements that have been done, thanks to the new {\it Fermi}-LAT catalogs, to solve this mystery. We demonstrate that the $\gamma$-ray intensity and spectrum of the IGRB is fully consistent with the unresolved emission from extragalactic point sources, namely Active Galactic Nuclei and Star Forming Galaxies. We show also that the IGRB can be employed to derive sever constraints for the $\gamma$-ray emission from diffuse processes such as annihilation of Dark Matter (DM) particles. Our method is able to provide low bounds for the thermal annihilation cross section for a wide range of DM masses.
\end{abstract}

\section{Introduction}
The existence of an isotropic component in the $\gamma$-ray sky has been observed since 1972 by the OSO-3 satellite \cite{1972ApJ...177..341K}. Recently the Large Area Telescope (LAT) on board the {\it Fermi} Gamma-Ray Space Telescope has provided a precise spectral measurement of this isotropic residual, called Isotropic diffuse $\gamma$-ray emission (IGRB), from 100 MeV to 820 GeV and at Galactic latitude ($|b|$) larger than $20^{\circ}$ \cite{Ackermann:2014usa}. The {\it Fermi}-LAT has also released the measurement of the Extragalactic $\gamma$-ray background (EGB), i.e. the superposition of the IGRB and the flux from detected sources.

The origin of the IGRB is one of the most intriguing mystery in astrophysics. Since the LAT has detected more than 3000 sources in four years of operation \cite{2015ApJS..218...23A}, the most obvious explanation for the solution of this puzzle is that the IGRB arises from the emission of unresolved, i.e. undetected, point sources.  Since Active Galactic Nuclei (AGNs), Star Formign Galaxies (SFGs) and pulsars are detected by {\it Fermi}-LAT at $|b| > 20^{\circ}$, we expect that these source populations should largely contribute to the IGRB intensity.

Blazars are AGNs with the jets aligned with the line of sight (LoS) and they costitute the most numerous source population in {\it Fermi}-LAT catalogs (see e.g. \cite{2015ApJS..218...23A}). They are therefore expected to give a large contribution to the IGRB between 10\% and 50\% (see e.g. \cite{DiMauro:2013zfa,2012ApJ...751..108A,Ajello:2013lka}).
According to the presence or absence of strong broad emission lines in their optical/UV spectrum, blazars are traditionally divided into flat-spectrum radio quasars (FSRQs) and BL Lacertae (BL Lacs) respectively \cite{1995ApJ...444..567P}.
Blazars are also classified using the frequency of the synchrotron peak into: low-synchrotron-peaked (LSP), intermediate-synchrotron-peaked (ISP) or as high-synchrotron-peaked (HSP) blazars.
FSRQs are almost all LSP blazars and they only contribute below 10 GeV to the IGRB explaining about 8\%-11\% of its intensity \cite{2012ApJ...751..108A}.
BL Lacs on the other hand are equally divided into LISP (LSP+ISP) and HSP sources and they contribute with about 10\% to the low energy part of the IGRB \cite{DiMauro:2013zfa,Ajello:2013lka} while they explain almost entirely the IGRB at energy larger than 100 GeV \cite{DiMauro:2013zfa}.

AGN with jets misaligned with respect to the LoS (MAGN) have weaker luminosities due to Doppler boosting effects \cite{1995ApJ...444..567P} and in fact only a dozen of them have been detected by the LAT \cite{2015ApJS..218...23A}. However using the existence of a correlation between the $\gamma$-ray and radio luminosities, it has been estimated that the unresolved population of MAGNs is very numerous hence they should explain a large fraction of the IGRB in the range $20$-$100\%$ as derived in Refs.~\cite{2011ApJ...733...66I,DiMauro:2013xta}.

SFGs are objects where the star formation rate is intense and $\gamma$ rays are produced from the interaction of cosmic rays with the interstellar medium or interstellar radiation field.
SFG $\gamma$-ray emission is very dim and the LAT has detected so far only nine individual galaxies \cite{2012ApJ...755..164A}. 
Many more SFGs have been detected at infrared and radio wavelengths therefore a numerous population of unresolved SFGs should exist and contribute between $4\%$ and $23\%$ of the IGRB as found in Ref.~\cite{2012ApJ...755..164A} using correlations between the infrared or radio band with $\gamma$-ray emission.

Finally, millisecond pulsars are the most promising Galactic contributors because many of these objects have been detected by the LAT at $|b|>10^{\circ}$. However in Ref.~\cite{Calore:2014oga} they have recently estimated that this source population can at most explain a few \% of the IGRB.

\begin{figure*}
\centering
\includegraphics[width=0.49\columnwidth]{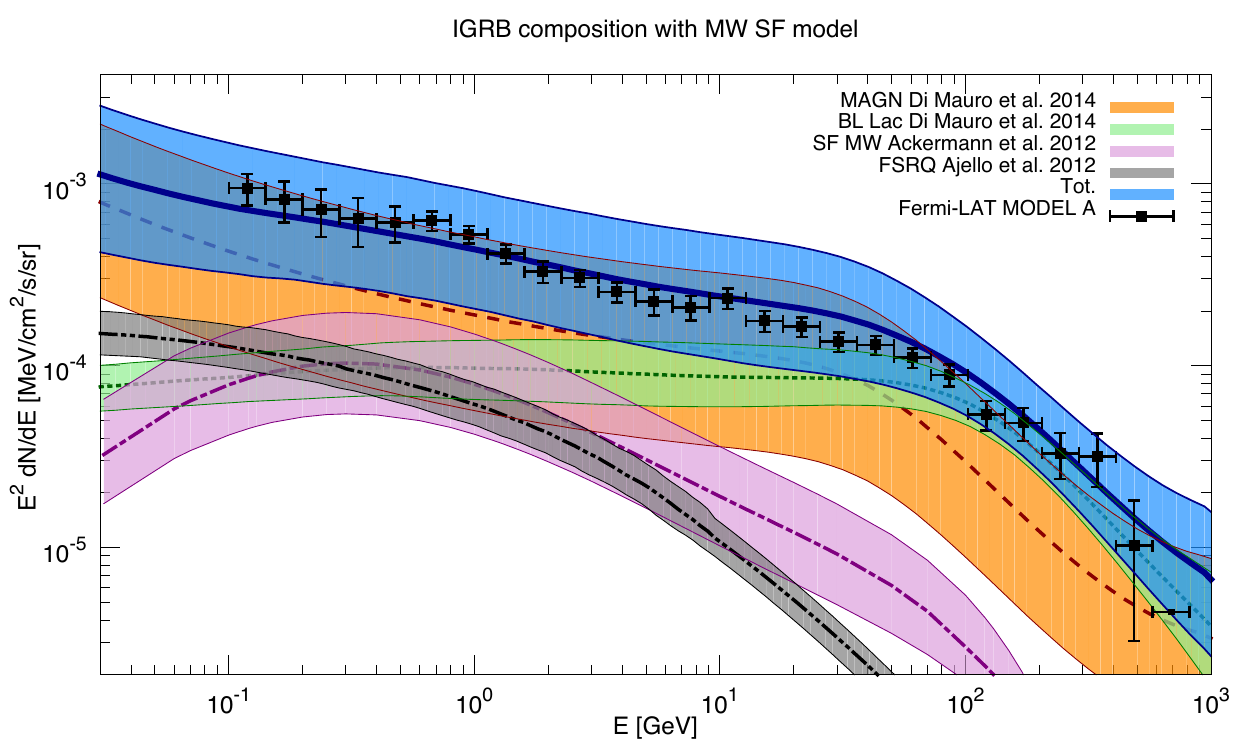}
\includegraphics[width=0.49\columnwidth]{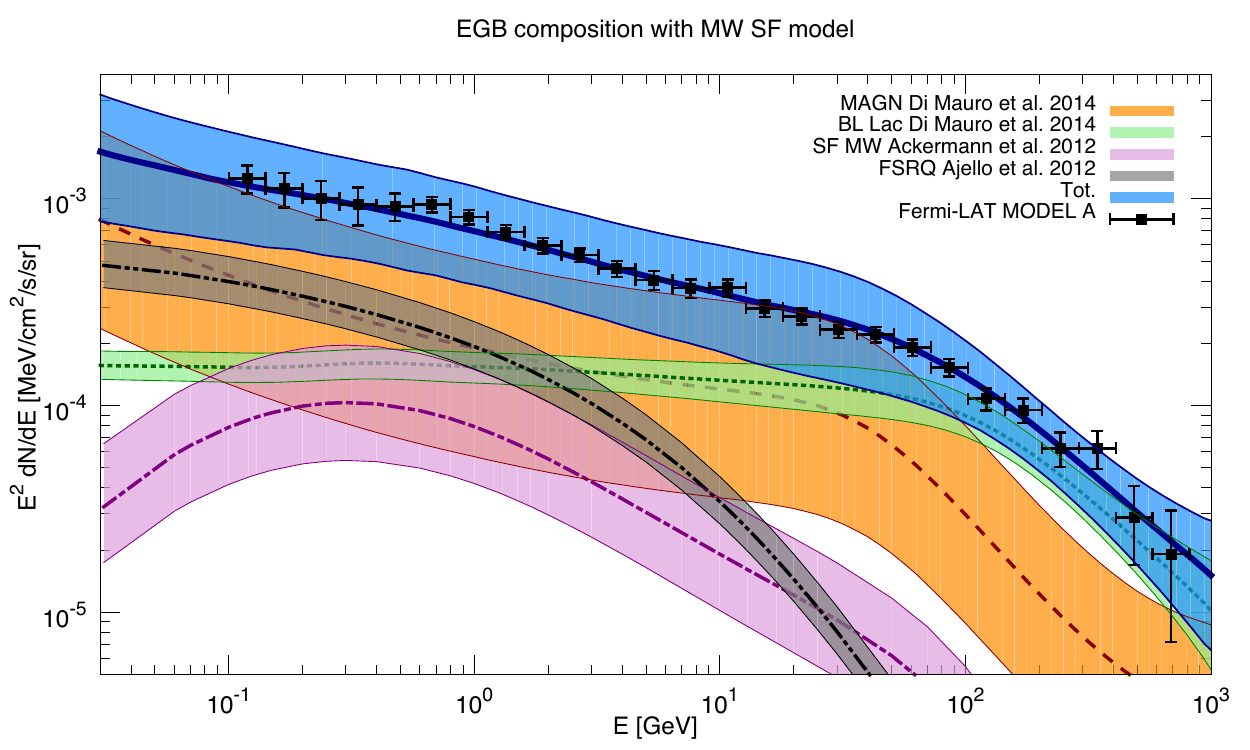}
\caption{Left (right) panel: average and uncertainty band for the $\gamma$-ray emission from unresolved (total=unresolved+resolved) MAGN, FSRQ, BL LAC and SFG sources, along with data for the IGRB (EGB) \cite{Ackermann:2014usa}.}
\label{fig:igrb_egb_theo}
\medskip
\end{figure*}

Unresolved FSRQ, BL Lac blazars together with MAGN and SFG sources have been showed to give a possible explanation of the IGRB and EGB intensity and spectrum in the entire energy range \cite{DiMauro:2013zfa,Ajello:2015mfa,DiMauro:2015tfa}
In Fig.~\ref{fig:igrb_egb_theo} for example we show the contribution  of the above cited extragalactic populations (MAGN \cite{DiMauro:2013xta}, FSRQ  \cite{2012ApJ...751..108A}, BL Lac \cite{DiMauro:2013xta} and SFG sources \cite{2012ApJ...755..164A}) to the {\it Fermi}-LAT data derived with Galactic foreground model (GFM) A.
In the rest of the paper we are going to use the same source populations and references reported in Fig.~\ref{fig:igrb_egb_theo}.

\section{The {\it Fermi}-LAT IGRB and EGB explained with emission from AGNs and SFGs}
\label{sec:astrofit}
In Ref.~\cite{DiMauro:2015tfa} the compatibility of the $\gamma$-ray emission from extragalactic sources to {\it Fermi}-LAT IGRB and EGB data has been tested on a statistical point of view.
In order to do so a $\chi^2$ analysis has been performed, including both the error of the EGB and IGRB data \cite{Ackermann:2014usa} and the uncertainties on the theoretical modeling of $\gamma$-ray emission from extragalactic sources.
We refer to \cite{DiMauro:2015tfa} for the full details of the analysis.
Two different approaches have been considered to account for the theoretical uncertainties. As a first case they have considered a renormalization of the average contribution with a 1-$\sigma$ error equal to the width of the bands showed in Fig.~\ref{fig:igrb_egb_theo}.
In the second and more realistic case changes in the spectral shapes of SFG and AGN emission are also included. For SFGs they have taken into account both the Power Law (PL) and Milky Way (MW) models (see \cite{2012ApJ...755..164A}) while for blazars they varied the photon index within the 1-$\sigma$ errors reported in the 2FGL catalog \cite{2012ApJS..199...31N}.

\begin{figure*}
\centering
\includegraphics[width=0.49\columnwidth]{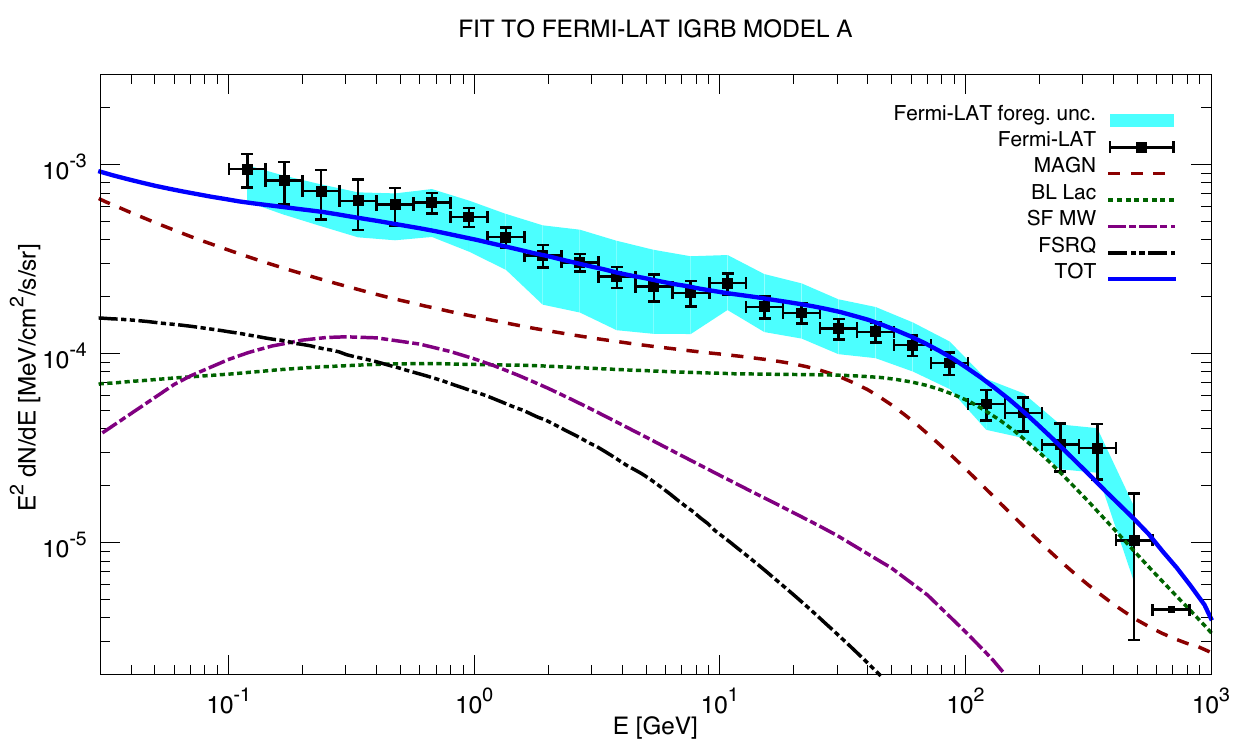}
\includegraphics[width=0.49\columnwidth]{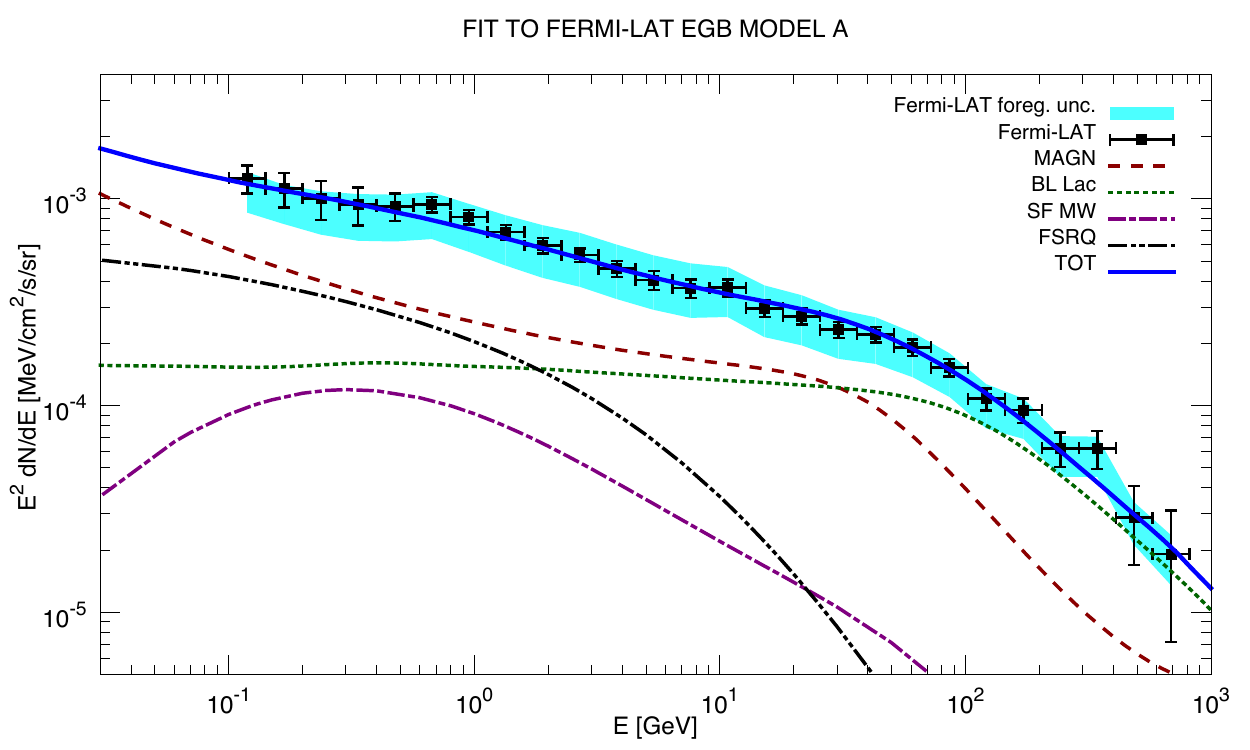}
\caption{The best-fit for the AGN and SFG $\gamma$-ray flux with GFM Model A is showed together with the systematic uncertainty of {\it Fermi}-LAT data (cyan band) \cite{Ackermann:2014usa}.}
\label{fig:backfit}
\medskip
\end{figure*}

The fits provide on average reduced $\chi^2$ ($\tilde{\chi}^2 = \chi^2/\rm{d.o.f.}$) values equal and also lower than 1 meaning that the $\gamma$-ray emission from AGN and SFG sources provide a good explanation of {\it Fermi}-LAT data.
The goodness of the fit depends on the GFM model employed to derive {\it Fermi}-LAT data and on average they derived better fits with Model C and with MW model of SFGs. The goodness of the fit is lower using GFM B and with PL spectrum for SFGs. In this case $\tilde{\chi}^2 \sim 2$.
We show in Fig.~\ref{fig:backfit} the best fit configuration.

\section{Resolving the {\it Fermi}-LAT EGB at $E>50$ GeV}
The {\it Fermi}-LAT Collaboration has recently released a new event-level analysis called Pass~8 \cite{Atwood:2013dra} that increased acceptance and improved point-spread function. Pass~8 permitted the creation of an all-sky survey at $>$50\,GeV with 360 $\gamma$-ray sources that constitutes the second catalog of hard {\it Fermi}-LAT sources (2FHL) \cite{Ackermann:2015uya}.

\begin{figure}[t]
	\centering
	\includegraphics[width=0.45\columnwidth]{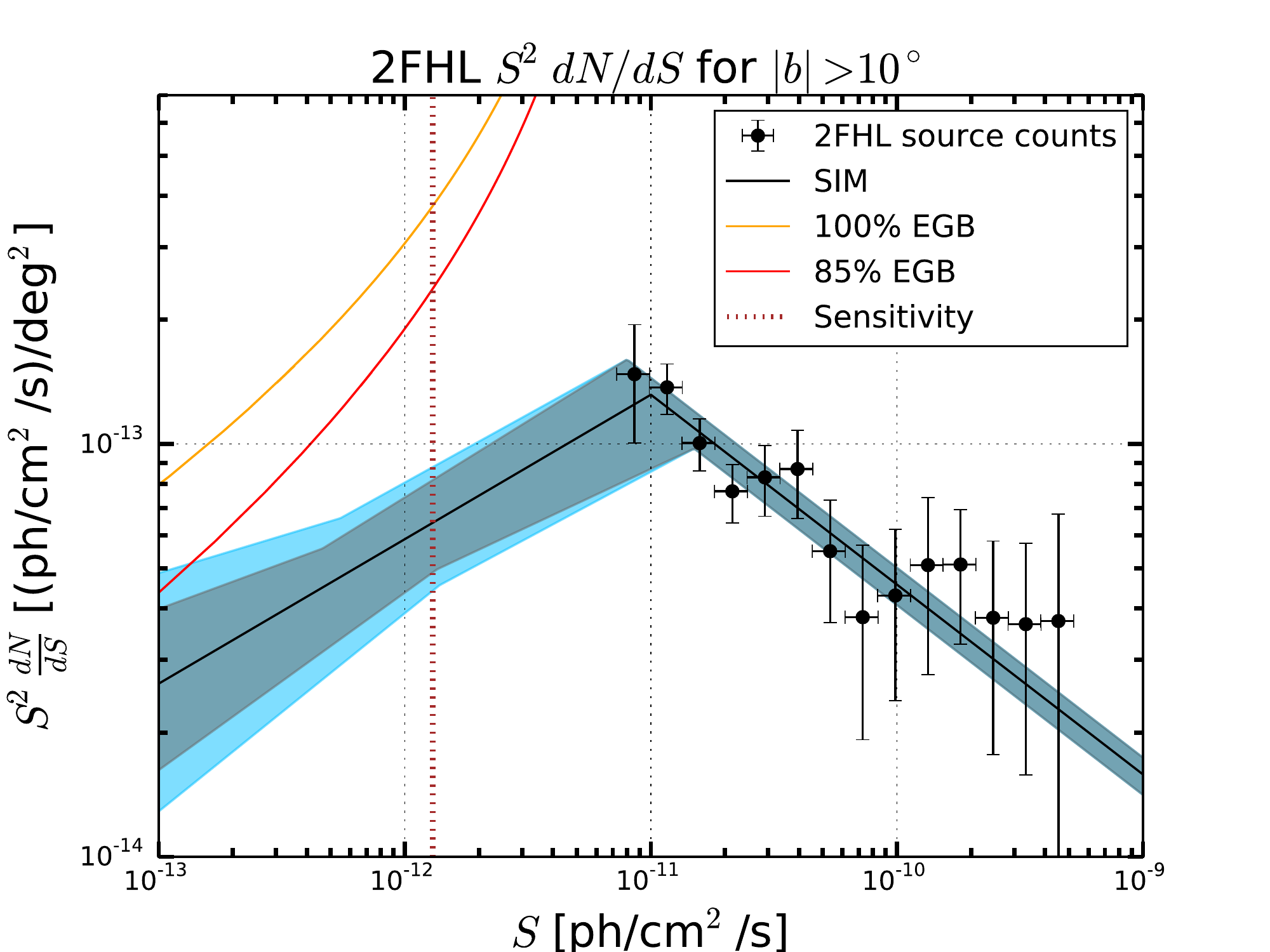}
	\includegraphics[width=0.45\columnwidth]{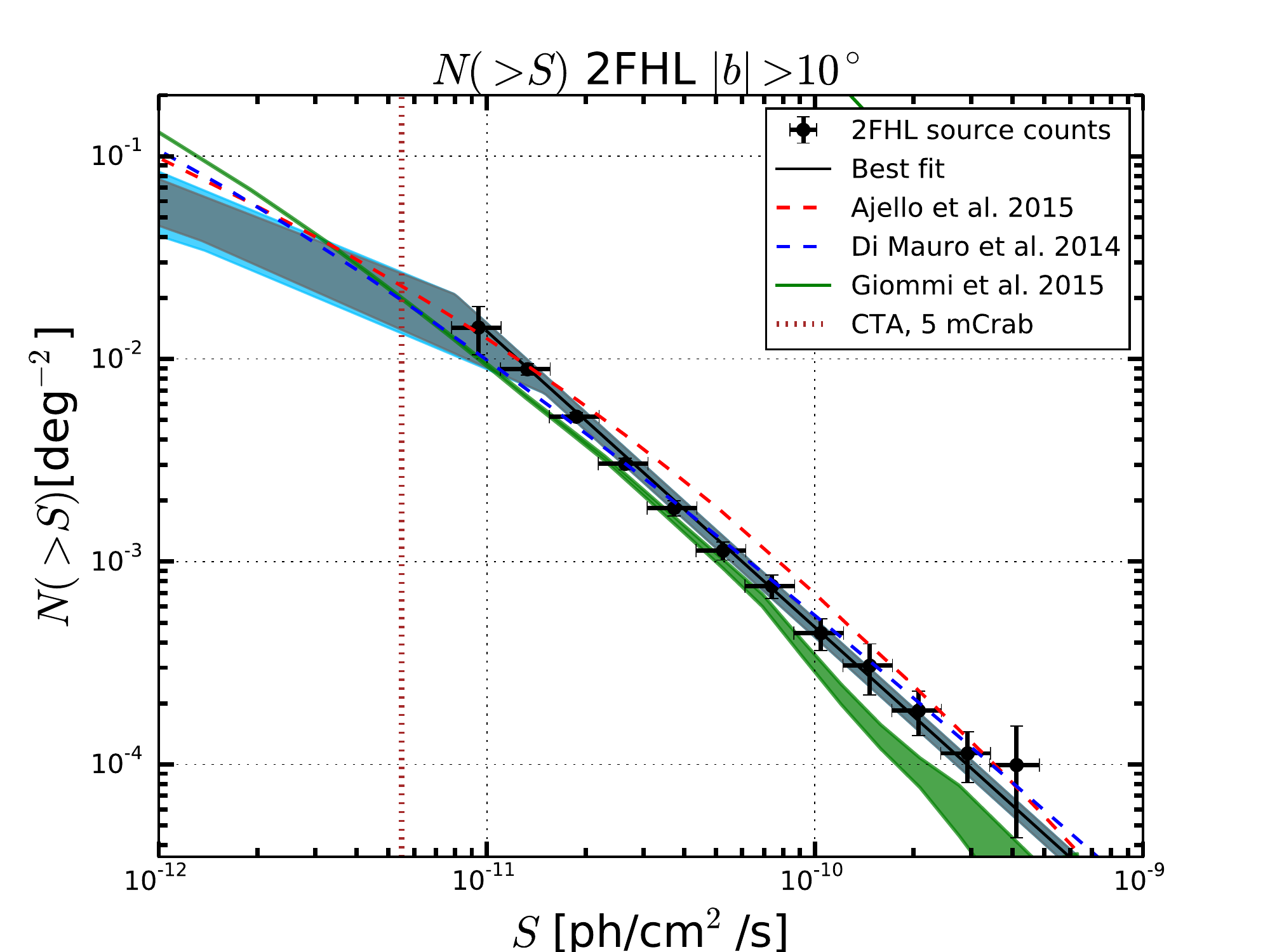}
\caption{Left Panel: best-fit ($1\sigma$ and $3\sigma$ uncertainty) for the intrinsic $S^2 dN/dS$ with black points (grey and cyan bands). The sensitivity of the photon fluctuation analysis (brown dotted line) and 85\% and 100\% level of the EGB intensity above 50 GeV (orange and red curves) are also displayed.
Right Panel: average (uncertainty band) of the measured cumulative source count distribution $N(>S)$ together with the theoretical predictions from Ref.~\cite{DiMauro:2013zfa} (blue dashed line), \cite{Ajello:2015mfa} (red dashed line) and \cite{2015MNRAS.450.2404G} (green band). The vertical dotted brown line shows the 5\,mCrab flux reachable by CTA in 240 hrs of exposure \cite{2013APh....43..317D}.}
\label{fig:2fhl} 
\end{figure}

The 2FHL is at $|b|>10^{\circ}$ dominated by blazars and in particular by BL Lacs (93\% BL Lacs and 4\% FSRQs).
In Ref.~\cite{TheFermi-LAT:2015ykq}, the source detection efficiency of the 2FHL catalog has been calculated using accurate Monte Carlo simulations of the $\gamma$-ray sky. A one-point photon fluctuation analysis has been also employed to derive the shape of the source count distribution $dN/dS$ up to fluxes a factor of about 10 lower than the source detection threshold. The derived $dN/dS$ (see left panel of Fig.~\ref{fig:2fhl}) is a broken power-law with an Euclidean slope (5/2) above the flux break $S_b$, located at $0.8-1.5\times10^{-11}$ ph\,cm$^{-2}$\,s$^{-1}$, and $1.60-1.75$ below $S_b$.
These results enabled the {\it Fermi}-LAT Collaboration to infer the intrinsic $dN/dS$ of sources located at $|b| > 10^{\circ}$. The presence of a break is measured by one-point photon fluctuation analysis but is also expected by the level of the EGB. If an Euclidean shape of the $dN/dS$ continued below the flux break, point sources would overshoot the 100\% contribution of the EGB below $1\times10^{-12}$ ph\,cm$^{-2}$\,s$^{-1}$ (see left panel of Fig.~\ref{fig:2fhl}).

Integrating the inferred flux distribution, we calculated that point sources and in particular blazars explain $86^{+16}_{-14}\%$ of the EGB. This result leaves small space to other source populations (such as SFGs) or diffuse processes (as $\gamma$-rays from annihilation or decay of Dark Matter (DM) particles \cite{Calore:2013yia,Fornasa:2015qua} or ultra high-energy cosmic ray interaction with the extragalactic background light \cite{2011PhRvD..84h5019A}) to contribute largely to the EGB.

\section{Dark matter constraints} 
$\gamma$ rays can be produced also from annihilation or decay of DM particles present in the Galactic halo or Galactic and extragalactic sub-halos.
We show in this section the upper limits for the DM annihilation cross section \sigmav derived in Ref~\cite{DiMauro:2015tfa} with the same $\chi^2$ method employed in Sec.~\ref{sec:astrofit}. They considered $\gamma$ rays produced from the main halo of our Galaxy directly with the so-called {\it prompt} emission and indirectly from inverse Compton scattering of electrons and positrons produced from DM annihilation off the interstellar radiation field.
The particle spectra from DM annihilation have been calculated with Pythia Montecarlo code \cite{Sjostrand:2007gs} for DM annihilations into $e^+e^-$, $\mu^+\mu^-$, $\tau^+\tau^-$, $b\bar{b}$, $t\bar{t}$ and $W^+W^-$ channels.
They have assumed an Einasto DM profile with a local DM density of $\rho = 0.4$ GeV/cm$^3$. 
We refer to \cite{DiMauro:2015tfa} for the full details of analysis.

\begin{figure*}
\centering
\includegraphics[width=0.49\columnwidth]{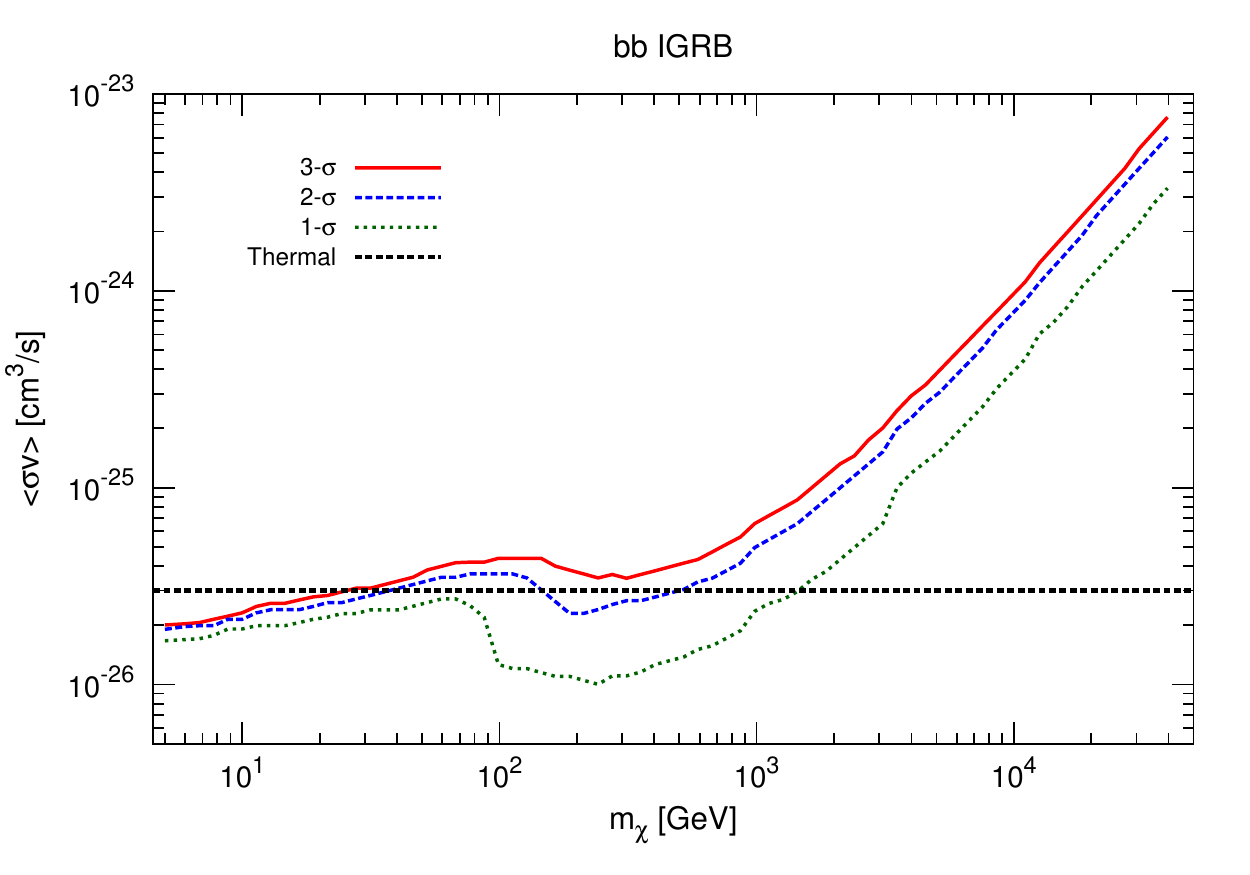}
\includegraphics[width=0.49\columnwidth]{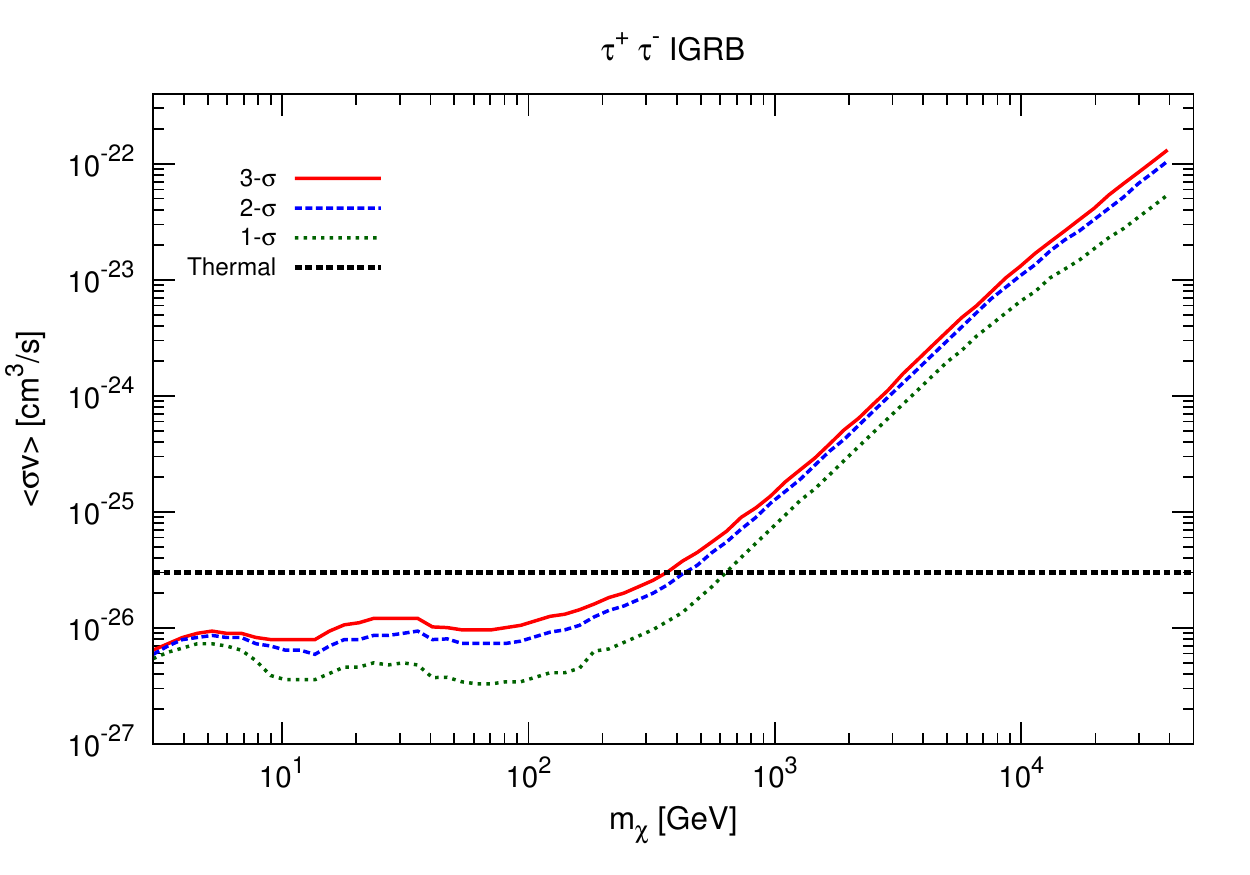}
\caption{3-$\sigma$, 2-$\sigma$ and 1-$\sigma$ C.L.s upper limits on the DM annihilation cross section \sigmav as a function of the DM mass $m_\chi$ for $\tau^+\tau^-$  and $b\bar{b}$ channels. The dotted horizontal line
indicates the thermal relic  annihilation value $\langle\sigma v\rangle=3\cdot 10^{-26} {\rm cm}^3 / {\rm s}$.}
\label{fig:UL_IGRB}
\medskip
\end{figure*}

In Fig.~\ref{fig:UL_IGRB} we show the results, derived with GFM A of IGRB data, for $\tau^+\tau^-$ and $b\bar{b}$ channels.
The limits constrain the thermal relic value for a wide range of DM masses and are competitive with bounds found with other analysis (e.g. \cite{Ajello:2015mfa}).
In Ref.~\cite{DiMauro:2015tfa} they displayed also how the upper limits change significantly considering different GFMs.

\begin{figure*}
\centering
\includegraphics[width=0.49\columnwidth]{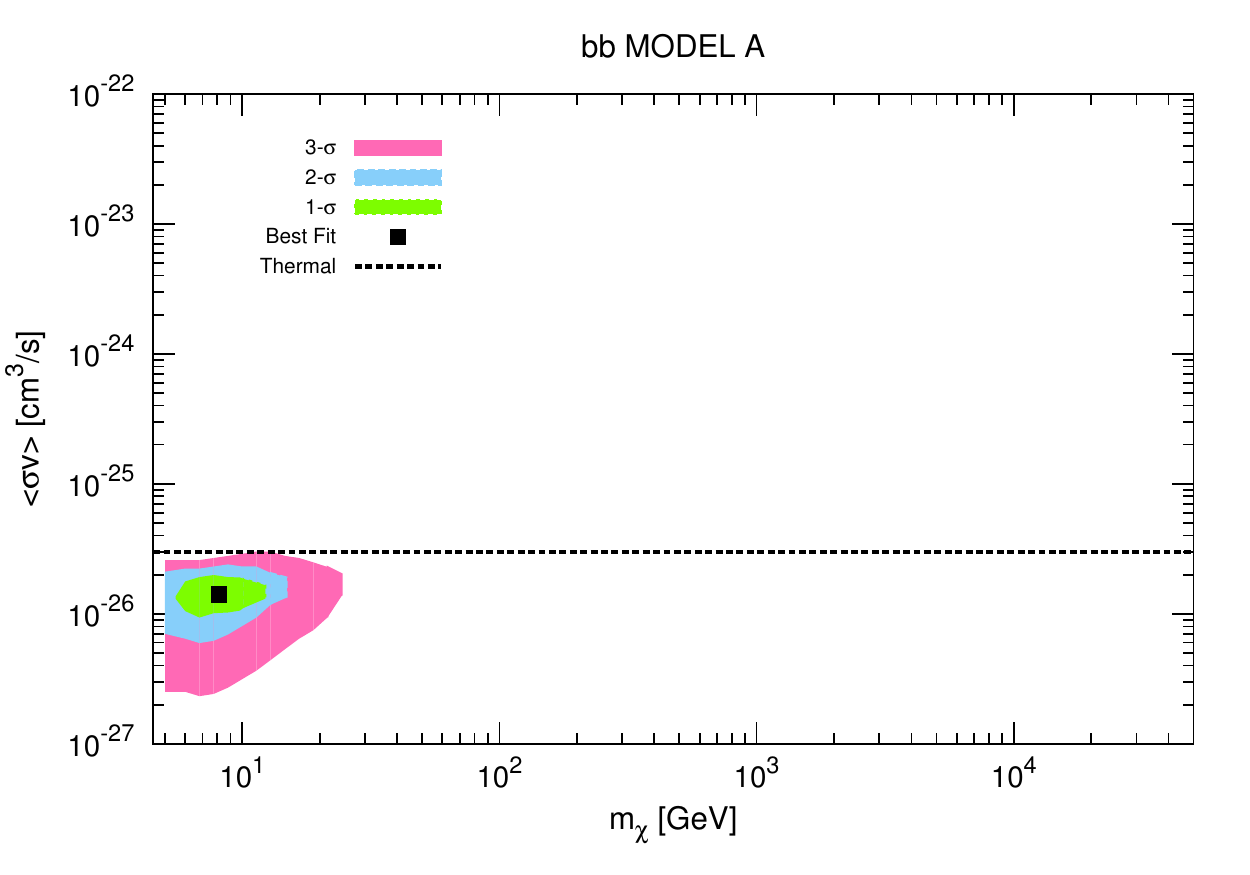}
\includegraphics[width=0.49\columnwidth]{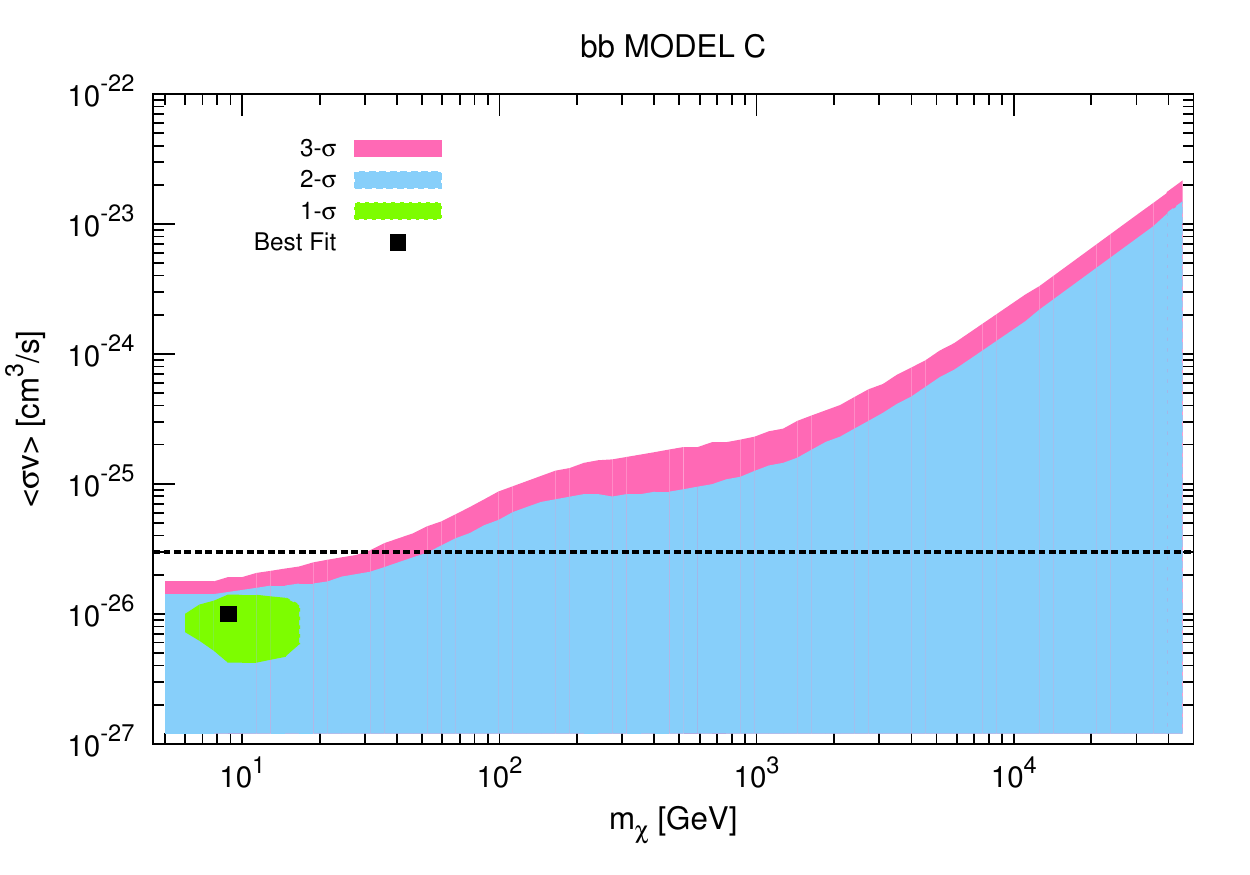}
\caption{1, 2 and 3-$\sigma$ C.L. contour plots in the $\langle\sigma v\rangle$-$m_\chi$ plane obtained from the fit to the IGRB data, derived with the GFM model A (left panel) and C (right panel), with the astrophysical backgrounds and a DM component.}
\label{fig:cp_igrb}
\medskip
\end{figure*}

They have also tested the possibility to find a DM signature in the IGRB spectrum. For this scope they have searched configurations where the addition of $\gamma$-ray emission from DM particles significantly improve the fit to {\it Fermi}-LAT data with respect to the case where only the $\gamma$-ray flux from point sources has been employed. Their best-fit case is for $b\bar{b}$ channel with a DM mass $8.2 \pm 2.3$ GeV and \sigmav $= 1.4 \pm 0.3$ cm$^3$/s (see left panel of Fig.~\ref{fig:cp_igrb}). This configuration is preferred to the case where only point sources contribute to the IGRB with a significance of about $3.5\sigma$.
However, the {\it Fermi}-LAT data are dominated by the systematic uncertainties due to the choice of GFM model as clearly visible by the width of the cyan band in Fig.~\ref{fig:backfit}.
In fact, considering the $b\bar{b}$ channel and GFM C, the DM contribution becomes much less significative and they can only set upper limits up to $3\sigma$ confidence level (see right panel of Fig.~\ref{fig:cp_igrb}).

%%%%%%%%%%%%%%%%%%%

\section{Conclusions}
In this paper we have showed that the last {\it Fermi}-LAT measurements permitted to start giving a solution to the puzzle of the IGRB origin. With a precise statistical analysis, in Ref.~\cite{DiMauro:2015tfa} they have demonstrated that AGN and SFG sources provide a viable explanation to both IGRB and EGB data. A detailed {\it Fermi}-LAT analysis focused at $E>50$ GeV confirmed this hypothesis finding that the EGB is almost all explained by blazars which dominate the 2FHL catalog.
Finally we have showed that the IGRB spectrum is very promising to set sever bounds for a DM contribution to the $\gamma$-ray sky and that this observable would be very sensitive to infer a DM signature if the systematic uncertainties will be significantly reduce in the future data release.

{\footnotesize
\section*{Acknowledgments}
The \textit{Fermi}-LAT Collaboration acknowledges support for LAT development, operation and data analysis from NASA and DOE (United States), CEA/Irfu and IN2P3/CNRS (France), ASI and INFN (Italy), MEXT, KEK, and JAXA (Japan), and the K.A.~Wallenberg Foundation, the Swedish Research Council and the National Space Board (Sweden). Science analysis support in the operations phase from INAF (Italy) and CNES (France) is also gratefully acknowledged.
}

\section*{References}
\bibliographystyle{iopart-num}
\bibliography{taup_short}

\providecommand{\newblock}{}
\begin{thebibliography}{10}
\expandafter\ifx\csname url\endcsname\relax
  \def\url#1{{\tt #1}}\fi
\expandafter\ifx\csname urlprefix\endcsname\relax\def\urlprefix{URL }\fi
\providecommand{\eprint}[2][]{\url{#2}}

\bibitem{1972ApJ...177..341K}
{Kraushaar} W~L {\em et~al.\/} 1972 {\em \apj\/} {\bf 177} 341

\bibitem{Ackermann:2014usa}
Ackermann M {\em et~al.\/} (Fermi-LAT) 2015 {\em \apj\/} {\bf 799} 86

\bibitem{2015ApJS..218...23A}
{Acero} F {\em et~al.\/} 2015 {\em \apjs\/} {\bf 218} 23

\bibitem{DiMauro:2013zfa}
Di~Mauro M {\em et~al.\/} 2014 {\em \apj\/} {\bf 786} 129

\bibitem{2012ApJ...751..108A}
{Ajello} M {\em et~al.\/} 2012 {\em \apj\/} {\bf 751} 108

\bibitem{Ajello:2013lka}
Ajello M {\em et~al.\/} 2014 {\em \apj\/} {\bf 780} 73

\bibitem{1995ApJ...444..567P}
{Padovani} P and {Giommi} P 1995 {\em \apj\/} {\bf 444} 567--581

\bibitem{2011ApJ...733...66I}
{Inoue} Y 2011 {\em \apj\/} {\bf 733} 66

\bibitem{DiMauro:2013xta}
Di~Mauro M {\em et~al.\/} 2014 {\em \apj\/} {\bf 780} 161

\bibitem{2012ApJ...755..164A}
{Ackermann} M {\em et~al.\/} 2012 {\em \apj\/} {\bf 755} 164

\bibitem{Calore:2014oga}
Calore F {\em et~al.\/} 2014 {\em \apj\/} {\bf 796} 1

\bibitem{Ajello:2015mfa}
Ajello M {\em et~al.\/} 2015 {\em \apj\/} {\bf 800} L27

\bibitem{DiMauro:2015tfa}
Di~Mauro M and Donato F 2015 {\em Phys.Rev.\/} {\bf D91} 123001

\bibitem{2012ApJS..199...31N}
{Nolan} P~L {\em et~al.\/} 2012 {\em \apjs\/} {\bf 199} 31

\bibitem{Atwood:2013dra}
Atwood W {\em et~al.\/} 2013 {\em \apj\/} {\bf 774} 76

\bibitem{Ackermann:2015uya}
Ackermann M {\em et~al.\/} (Fermi-LAT) 2015 {\em arXiv:1508.04449\/}

\bibitem{TheFermi-LAT:2015ykq}
Ackermann M {\em et~al.\/} 2015 {\em arXiv:1511.00693\/}

\bibitem{Calore:2013yia}
Bringmann T {\em et~al.\/} 2014 {\em Phys. Rev.\/} {\bf D89} 023012

\bibitem{Fornasa:2015qua}
Fornasa M and S\'anchez-Conde M~A 2015 {\em arXiv:1502.02866\/}

\bibitem{2011PhRvD..84h5019A}
{Ahlers} M and {Salvado} J 2011 {\em \prd\/} {\bf 84} 085019

\bibitem{2015MNRAS.450.2404G}
{Giommi} P and {Padovani} P 2015 {\em \mnras\/} {\bf 450} 2404--2409

\bibitem{2013APh....43..317D}
{Dubus} G {\em et~al.\/} 2013 {\em Astroparticle Physics\/} {\bf 43} 317--330

\bibitem{Sjostrand:2007gs}
Sjostrand T, Mrenna S and Skands P~Z 2008 {\em Comput.Phys.Commun.\/} {\bf 178}
  852--867

\end{thebibliography}

\end{document}